\documentclass[prb,aps,showpacs,floats,twocolumn]{revtex4}
\usepackage{graphicx}
\usepackage{amsmath}
\usepackage{float}
\usepackage{bbold}
\usepackage[titletoc,toc,title]{appendix}
\usepackage{color}

\newcommand{\eeqref}[1]{Eq.~\ref{#1}}

\begin{document}

\title{Non-equilibrium variational cluster perturbation theory:\\
quench dynamics of
 the quantum Ising model}
\author{Mohammad Zhian Asadzadeh$^{1,2}$, Michele Fabrizio$^2$, and Enrico Arrigoni$^1$ }
\affiliation{ 
$^1$ Institute of Theoretical and Computational Physics, Graz University of Technology, 8010 Graz, Austria\\
$^2$ International School for Advanced Studies (SISSA), Via Bonomea 265, 34136 Trieste, Italy
}

\begin{abstract}
We introduce
 a variational implementation of cluster perturbation theory (CPT) 
to address the dynamics  of 
spin systems driven out of equilibrium. We benchmark the method with the 
quantum Ising model subject to a sudden quench of the transverse magnetic field across the transition or within a phase.
We treat both the one-dimensional case, for which an exact solution is available,
as well 
 the  two-dimensional one, 
for which has to resort to numerical results.
 Comparison with exact results shows that the approach provides a quite accurate description of the real-time dynamics up to a characteristic time scale $\tau$ that increses with the size of the cluster used for CPT. In addition, and not surprisingly $\tau$ is small for quenches across the equilibrium phase transition, but can be quite larger for quenches within the ordered or disordered phases. 
\end{abstract}


\maketitle
\begin{section}{Introduction}
The remarkable progresses  
of experiments on ultracold atoms trapped in optical lattices\cite{Bloch,Lewenstein,Strohmaier} have boosted  a great interest in the nonequilibrium dynamics of closed quantum systems, especially when they are suddenly pushed across a quantum critical point\cite{Sachdev,Suzuki}. A rich theoretical activity 
thus flourished, starting from the paradigmatic example of quantum criticality, namely the  quantum Ising model.\cite{Meinert,Rossini,Calabrese}

Developing suitable tools for handling many-body systems out of equilibrium is a big challenge  that started some time ago with the pioneering works by Kubo \cite{Kubo}, Schwinger \cite{Schwinger}, Kadanoff and Baym \cite{Kadanoff}, and Keldysh \cite{Keldysh}. This effort continued with the work by Wagner\cite{Wagner},  who unified the Feynman, Matsubara, and Keldysh perturbation theories into a single and very flexible formalism, till latest developments related to dynamical mean field and related cluster-embedding methods (see, e.g. \cite{Freericks,Schmidt,Hofmann,Jung,Knap3,Aoki,Balzer1,Balzer2,Gramsch,ar.kn.13,do.nu.14,do.ga.15}). We shall in particular be concerned with the very recent out-of-equilibrium generalization of cluster perturbation 
theory (CPT) \cite{Senechal,Gros},  which is attractive and conceptually simple \cite{Balzer2}.  In CPT  the  lattice is 
divided into small clusters which can be diagonalized exactly. 
The inter-cluster terms are then treated within strong-coupling perturbation theory. Its nonequilibrium version allows to
investigate the unitary quantum evolution in the thermodynamic limit, accounting for non-local correlations on a length scale defined by the size of the considered cluster. Besides the simplicity of the formulation, the efficiency and accuracy of the specific implementation is also of major  
importance.

The main purpose of this work is to develop a non-equilibrium variational implementation of CPT for spin systems. We test the method on the quantum Ising model after a sudden quench of the transverse field.  Since the model is exactly solvable in one dimension we have the possibility to benchmark the approach. We also investigate the same model in two dimensions where an exact solution is not available. In this case, we compare  with finite-size exact diagonalization results. We discuss in detail how to efficiently  
implement the method so to allow reaching relatively long simulation times with moderate computational effort. 

The paper is organised as follows. In section~\ref{NE-GF} the non-equilibrium Green's function formalism is briefly presented. The model we shall study is introduced in section~\ref{Hamiltonian}. 
Section~\ref{Method} describes the CPT method together with its self-consistent variational improvement. Results are reported  in Sec.~\ref{Results}. Section~\ref{Summary} is devoted to concluding remarks.

\end{section}

\begin{section}{Non-Equilibrium Green's functions} \label{NE-GF}
In this section we briefly outline the non-equilibrium Green's function formalism to set up the notations that we shall use throughout the paper. There is 
a wide literature on the subject but in this work we mainly follow the Kadanoff-Baym-Wagner scheme \cite{Kadanoff,Wagner}.

Consider a system initially  (at time $t_0=0$) at equilibrium described by a  Hamiltonian $H_{eq}$ and temperature $1/\beta$. At $t>t_0$ a 
generic time dependent Hamiltonian $H(t)$ is switched on. The non-equilibrium formalism works through averages of time-ordered products of operators 
along the Kadanoff-Baym contour\cite{Kadanoff,Keldysh,Danielewicz,Rammer} shown in Fig. \ref{contour}. The contour is composed of three branches: it starts at $t_0=0$, runs up to $t_{max}$ and then back to the
initial time, and finally moves parallel to the imaginary axis up to $\tau=-i\beta$. \\
  \begin{figure}[t!]
    \centering
    \includegraphics [scale=0.15]{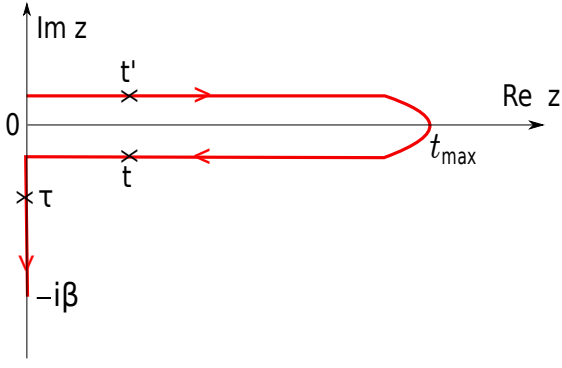}
    \caption{The $L$ shaped Kadanoff-Baym contour ${\cal C}$ . The arrows indicate the contour ordering. For example, $t'$ lies ahead of $t$ in the ordering ($t>t'$), i.e., operators at $t'$ 
             are sorted to the right by the contour ordering}
    \label{contour}
\end{figure}
Due to the lack of time translation invariance, 
the non-equilibrium single-particle Green's function depends on two time variables rather than on their difference and is defined as the contour ordered expectation value
 \begin{equation}
 \label{GF-definition}
 \begin{split}
  G_{i,j}&(z,z')=-i\langle {\cal T}_{\cal C} a_i(z) a_j^\dag(z') \rangle=\\
  &=-i\theta^{\cal C}(z-z')\langle a_i(z) a_j^\dag(z')\rangle -i\theta^{\cal C}(z'-z)\langle  a_j^\dag(z') a_i(z)\rangle
  \,,
  \end{split}
 \end{equation}
where $a_i^\dag(a_i)$ are the creation (annihilation) operators for particles, in the present case bosons, at site $i$ and $z,z'$ are variables on 
the contour ${\cal C}$, and can be real or imaginary depending on the branch of the contour in which they lay.
 The time evolution of 
the operators on the Kadanoff-baym contour is defined in the Heisenberg picture with Hamiltonian $H(z)$. ${\cal T}$ is the time ordering operator and 
is defined via the contour step function $\theta^{\cal C}(z-z')$. The averages in Eq. (\ref{GF-definition}) are over the initial 
equilibrium Hamiltonian $H_{eq}$ at temperature 
$1/\beta$. 

 The Dyson equation reads   
\begin{equation}
\label{Dyson}
 \hat{G}=\hat{G_0}+\hat{G_0}\bullet \hat{\Sigma}\bullet \hat{G}\,,
\end{equation}
where $\hat{G_0}$ is the bare Green's function and $\hat{\Sigma}$ the self-energy. The product symbol $\bullet$ denotes 
 the matrix multiplication in space and the integration 
over the time variables along the contour ${\cal C} $.

For a given Green's function $\hat{G}(z,z')$ each variable $z,z'$ can lay  on one of the three branches of the contour in Fig.~\ref{contour}. This prompts an alternative representation of
$\hat{G}$ as a $3\times 3$ matrix, as introduced by  Wagner \cite{Wagner}.  
Of the $9$ matrix elements, only $6$ are linearly independent, so that after a suitable transformation one is left with $6$ nonzero terms, which 
are referred to as 
 the retarded $(G^R)$, advanced $(G^A)$, Keldysh $(G^K)$, left-mixing $(G^\rceil)$, right-mixing $(G^\lceil)$ and 
Matsubara Green's function $(G^M)$. They are explicitly given as
\begin{equation}
\label{wagner-matrix}
\begin{split}
G_{i,j}^R(t,t')=-i\theta(t-t')\langle [a_i(t),a_j^\dag(t')]\rangle\,,\\
G_{i,j}^A(t,t')=G_{j,i}^R(t',t)^*\,,\\
G_{i,j}^K(t,t')=-i\langle \{a_i(t),a_j^\dag(t')\}\rangle \,,\\
G_{i,j}^\rceil(t,\tau)=-i\langle a_j^\dag(\tau)a_i(t)\rangle \,,\\
G_{i,j}^\lceil(\tau,t)=-i\langle a_i(\tau) a_j^\dag(t)\rangle \,,\\
G_{i,j}^M(\tau,\tau')=-\langle {\cal T}_\tau a_i(\tau) a_j(\tau') \rangle \,,
\end{split}
\end{equation}
where $t$ and $t'$ are real times and $\tau,\tau'\in [0, -i \beta]$.  In the above equations $\{\dots\}$ and $[\dots]$ stem for anticommutator and commutator, respectively.

\end{section}

\begin{section}{Hamiltonian}\label{Hamiltonian}
 The Hamiltonian of the Ising model in a transverse field is given by
\begin{equation}
\label{hamilt}
 H=-J\sum_{\langle i,j\rangle} S_i^x S_j^x+h\sum_i S_i^z\, ,
\end{equation}
where $\langle i,j\rangle$ means summation over nearest neighbor spins, and $h$ is the strength of the magnetic field, with $J>0$ and $h>0$. In the 
following we shall work in units of $J=1$. The Hamiltonian of Eq. (\ref{hamilt}) in one dimension has an exact solution which is obtained by a   
Jordan-Wigner transformation that maps the system onto a quadratic Hamiltonian for spinless fermions, which  can be exactly solved \cite{Lieb,Pfeuty}. On the other hand, in two dimensions  
an exact solution is not available \cite{Oleg}.

Cluster embedded techniques such as CPT in equilibrium have been applied to fermionic and bosonic systems\cite{Potthoff_vca1,Potthoff_vca2,Koller,Aichhorn,Zacher}. Out of 
equilibrium, CPT has been applied to the fermionic Hubbard model\cite{Balzer2,Gramsch}. Here we formulate nonequilibrium
 CPT for spin systems, exploiting the well known equivalence between spin-1/2 operators are hard-core bosons. We also provide a variational improvement of it, which allows to treat the ordered phase.\\

Specifically, if we assume that spin-up corresponds to the presence of a hard-core boson, and spin-down to its absence, the following relationships between spin and boson operators hold\cite{Sachdev}
\begin{equation}
 \begin{split}
  & S^+ \rightarrow a^\dag \,,\\
  & S^-  \rightarrow a \,,\\
  & S^z \rightarrow a^\dag a -1/2\,,\\
  & S^x \rightarrow (a+a^\dag)/2\,,
 \end{split}
\end{equation}
where $S^+$ and $S^-$ are raising and lowering spin operators, respectively. 

The Hamiltonian in the bosonic representation then becomes
 \begin{equation}\label{hamilt-boson}
 \begin{split}
  H=-\frac{J}{4}\sum_{\langle i,j\rangle} (a_i a_j+a_i^\dag a_j^\dag+a_i a_j^\dag+a_i^\dag a_j)\\
  +h\sum_i (a_i^\dag a_i-1/2)+U\sum_i n_i (n_i-1)
  \end{split}
 \end{equation}
where the on-site Hubbard-like term enforces the hard-core constraint when $U\to \infty$, and 
$n_i=a^\dag_i a_i$. Since the Hamiltonian 
contains both normal and anomalous hopping terms, Green's functions with  anomalous terms are needed to study the system (see appendix~\ref{cpt-numeric}).

\end{section}


\begin{section}{Method} \label{Method}
\begin{subsection}{Cluster perturbation theory}
Cluster perturbation theory (CPT)\cite{Gros,Senechal} is a simple quantum cluster method to deal with correlated systems. In this approach the idea is to embed a finite 
cluster of sites, for which a numerically exact solution is affordable, into the infinite lattice. In practice the starting point is to partition the original $D$-dimensional 
lattice of linear size $L$ into clusters of linear size $L_c$ with open boundaries. Fig. \ref{lattice} shows an example for a tiling in $D=1$ and $L_c=4$.
\begin{figure}[t!]
    \centering
    \includegraphics [scale=0.2]{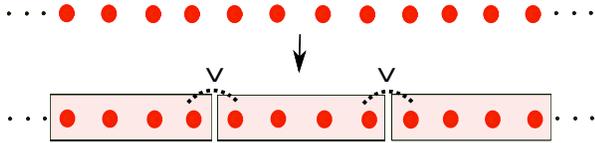}
    \caption{Partitioning of a D=1 lattice into clusters with size $L_c=4$. The inter-cluster hopping is denoted by $V$ }
    \label{lattice}
\end{figure}
All clusters are considered as supercells that form a superlattice, each supercell being identified by a superlattice vector $r$. The sites within each cluster 
are in turn labelled by vectors $R$. The lattice Hamiltonian $H$ is thus written as 
\begin{equation}\label{separat}
 H=H_0+V\,,
\end{equation}
where $H_0$ corresponds to the cluster Hamiltonian and $V$ describes the inter-cluster terms. CPT Green's function can be obtained by a subsequent expansion 
in powers of the 
inter-cluster hopping. {Diagrammatic \cite{Metzner,Pairlaut} and cluster dual fermion approaches\cite{Hafermann} provide a systematic  
expansion in terms of the inter-cluster terms, which then has to be 
truncated at some order}. Within strong-coupling 
perturbation theory\cite{Senechal,Senechal2} one obtains an expression for the lattice Green's function at lowest order 
\begin{equation}
G(\omega)=G_0(\omega)-G_0(\omega)VG(\omega)\,,
\end{equation}
where $V$ is the matrix representation of the inter-cluster hopping, $G_0(\omega)$ the exact equilibrium Green's function of the cluster and the product is just a matrix multiplication 
in lattice sites. The Green's function $G_0$ is diagonal in  $r$ and identical for all supercells, whereas $V$ is off-diagonal in $r$. Because of superlattice translation invariance, the above equation is simpler in momentum space. After partial Fourier transform, $r\rightarrow q$, the CPT equation transforms into 
\begin{equation}
\label{cpt-eq}
G(q,\omega)=G_0(\omega)-G_0(\omega)V(q)G(q,\omega)\,,
\end{equation}
where now $G$, $G_0$ and $V$ are matrices in the label $R$ of the sites within each supercell. The CPT is a conceptually 
simple method that nevertheless includes short-range correlations on the scale of cluster size and therefore requires moderate computational resources.\\

The idea of CPT  can be straightforwardly transferred to non-equilibrium situation by replacing the equilibrium frequency-dependent Green's functions with the contour 
ordered ones. The authors of Ref.~\cite{Balzer2} have developed a non-equilibrium formulation for CPT (NE-CPT) and  examined how the technique works for the Fermi-Hubbard model. 
The NE-CPT equation reads as following
\begin{equation}
\label{ne-cpt-equation}
\hat{G}(q)=\hat{G}_0+\hat{G_0}\bullet \hat{V}(q)\bullet \hat{G}(q)\,.
\end{equation}
The solution of Eq.~(\ref{ne-cpt-equation}) provides the non-equilibrium CPT Green's function $\hat{G}(q)$. In the NE-CPT equation the product symbol $\bullet$ denotes not only 
the matrix multiplication but also an integration over time variables along the contour $\cal {C}$. Furthermore $\hat{V}(q)=V(q)\bigotimes \mathbb{1}$ where
$\mathbb{1}$ is a $\delta$-function on the contour, i.e., $\delta(z'-z)=\mathbb{1}$. 
In what follows we shall omit  the momentum dependence to simplify notations. The explicit integral form of Eq.~(\ref{ne-cpt-equation}) is
\begin{equation}\label{contour-eq}
 \hat{G}(z,z')=\hat{G}_0(z,z')+\int_{\cal {C}} dz_1 \,\hat{G}_0(z,z_1)V\hat{G}(z_1,z')\,,
\end{equation}
where integration is carried out along the three branches of the contour $\cal{C}$ in Fig. \ref{contour}, i.e. 
\begin{equation}
 \int_{\cal {C}} dz_1=\int_0^{t_{max}} dt-\int_0^{t_{max}} dt+\int_0^{-i\beta} d\tau \;.
\end{equation}
The numerical solution of the generic contour equation (\ref{contour-eq}) requires discretization of the time variable. A straightforward 
but not efficient solution for $\hat{G}$ involves a matrix inversion\cite{Balzer2} where large matrices in discretized time are used. In this manner  
reaching long time dynamics is computationally prohibitive. {Alternatively, by using the Kadanoff-Baym equations \cite{Kadanoff,Bonitz} 
one can derive the same integral equation, Eq. (\ref{contour-eq}, for the  components of $\hat{G}$ in the Wagner representation (see Eq. (\ref{wagner-matrix})).
A practical application of this approach to non-equilibrium dynamical mean-field theory (NE-DMFT) has been 
presented by Tran\cite{Tran}.}
This method takes advantage of the causality of the integral equations: the properties of the system 
at specific time $t=t_1$ do not depend on the information at $t>t_1$ and so its {\sl a priori} knowledge is not required in the calculation. Here we follow this approach but for spatially inhomogeneous systems. For  
details of the procedure and technical issues see Appendix~(\ref {cpt-numeric}) 
\end{subsection}
\begin{subsection}{Variational cluster perturbation theory}
 Within CPT  one is free to add an arbitrary single-particle term $-\Delta$ to the cluster Hamiltonian $H_0$ (Eq. (\ref{separat})) provided that it is then
 subtracted perturbatively, i.e. added to $V$, such that the Hamiltonian $H$ remains unchanged. The CPT expansion is now carried out in the new 
 perturbation $\bar{V}=V+\Delta$ with the new cluster Hamiltonian $H'=H_0-\Delta$. While ideal exact results should not 
 depend on $\Delta$, in practice results do depend on $\Delta$ due to the approximate nature of 
 the CPT expansion. \\
 In this work we shall consider a $Z_2$ symmetry breaking term 
 \begin{equation}
\label{anomalous}
  \Delta=\sum_{R=1}^{L_c} f_R\,S_R^x= \sum_{R=1}^{L_c} \frac{f_R}{2}\Big(a_R+a_R^\dag \Big)\,,
 \end{equation}
where $f_R$ are real variational parameters to be fixed. We show below that accounting for this variational term is crucial to describe the 
ordered phase of quantum Ising model. The optimum value of the variational parameters $f_R$ should be 
determined through a variational principle\cite{Hofmann,Enrico,Knap2}. Here
we shall resort to a simplified version of the variational 
procedure introduced in ~\cite{Enrico,Knap2}. Specifically, we fix the variational parameters within a self-consistent approach where the inter-cluster term $S_i^x\,S_j^x$ is replaced with its mean-field approximation as 
\begin{equation}
 S_i^xS_j^x=\langle S_i^x \rangle S_j^x+S_i^x \langle S_j^x \rangle -\langle S_i^x \rangle \langle S_j^x \rangle\,.
\end{equation}
In one dimension, for example, upon tiling the infinite lattice into clusters of size $L_c$,  the mean-field expression for the supercell Hamiltonian at equilibrium is
\begin{eqnarray}
\label{hp}
 H'&=&-J\sum_{R=1}^{L_c-1} S_R^xS_{R+1}^x+h_0\sum_{R=1}^{L_c} S_R^z
 \nonumber\\
 && - f_{L_c}\,S_1^x - f_{1}\,S_{L_c}^x\,,
\end{eqnarray}
where $f_R = J\langle S_{R}^x \rangle$, $R=1,L_c$ are 
the mean-field self-consistency conditions 
and, by translational symmetry, we shall set $f_1=f_{L_c}$.\\

Out of equilibrium the variational parameters become time dependent. The protocol we shall implement is a sudden quench of the magnetic field from $h_0$ 
to a different value $h$. Therefore the explicit time dependent mean-field cluster Hamiltonian becomes
\begin{eqnarray}
 H'(t)&=&-J\sum_{R=1}^{L_c-1} S_R^xS_{R+1}^x+h\sum_{R=1}^{L_c} S_R^z
 \nonumber\\
 && - f_{L_c}(t)\,S_1^x - f_{1}(t)\,S_{L_c}^x\,,
\end{eqnarray}
with the self-consistency condition  
\begin{equation}
 f_R(t)=f_R(t)^*=J\,\langle \Psi(t)| S_{R}^x|\Psi(t)\rangle\,,
\end{equation}
where $|\Psi(t)\rangle$ 
is the time evolved 
 cluster wavefunction.  
In order to evaluate the time dependent variational parameters $ f_R(t)$ we expand 
the latter to linear order 
\begin{equation}
 |\Psi(t+\Delta t)\rangle \approx \Big(1-iH'(t)\Delta t\Big)\,|\Psi(t)\rangle+\mathcal{O} (\Delta t^2)
\end{equation}
starting from the initial 
equilibrium state $|\Psi(t=t_0)\rangle$. As a result, the parameters $f_i(t+\Delta t)$ 
can be taken as
\begin{equation}
 f_i(t+\Delta t)\approx J\langle \Psi(t+\Delta t)| S_i^x|\Psi(t+\Delta t)\rangle
\end{equation}
at each time step. 

\end{subsection}

\begin{subsection}{CPT corrections to the order parameter}
Due to the presence of anomalous terms linear in creation and annihilation operators the new perturbation $\bar{V}$ including $\Delta$ (\eeqref{anomalous})
is not quadratic in the boson operators and therefore one has to generlaize CPT to deal with anomalous terms as well-
The way to do this 
(see Ref.~\cite{Knap2,Enrico})
is
to first perform starndard CPT on top of  the cluster Hamiltonian $H'$ (\eeqref{hp}) 
by using just the quadratic part of
 $V$ as a perturbation. 
The CPT correction to the condensate can be then obtained 
by using
an expression
derived within a so-called pseudoparticle formulation of CPT~\cite{Knap2} 
for the  Bose-Hubbard model in the superfluid phase, and,
 subsequently confirmed more formally within a
 self-energy 
functional approach.~\cite{Knap2}
For the equilibrium case, one obtains
\begin{equation}\label{condens}
 G^{-1}\langle A\rangle=G^{'-1}\langle A \rangle'+F\,,
\end{equation}
where
 $G$ and $\langle A\rangle$ are the
CPT corrected Green's function and expectation value of the condensate, respectively, while the terms  
 with prime stands for their cluster values. The vector $F$ describes the variational parameters $f$ of 
 Eq.~(\ref{anomalous}).  In Eq.~(\ref{condens}) the Green's functions 
 are $2L_c\times 2L_c$ Nambu matrices and the  $\langle A\rangle$,  $\langle A \rangle'$ and $F$ are $2L_c$ Nambu vectors, namely
 \begin{equation}
 \langle A\rangle'=\begin{bmatrix}
     \langle a'_1 \rangle\\
     .  \\
     .\\
     .\\
     \langle a'_{L_c}\rangle \\
     \langle a_1^{' \dag} \rangle \\
     .\\
     .\\
     .\\
     \langle a_{L_c}^{' \dag} \rangle  
    \end{bmatrix}
    \;,\;
   \langle A\rangle =\begin{bmatrix}
     \langle a_1\rangle\\
     .  \\
     .\\
     .\\
     \langle a_{L_c}\rangle\\
     \langle a_1^\dag \rangle \\
     .\\
     .\\
     .\\
     \langle a_{L_c}^\dag \rangle
    \end{bmatrix}
\;,\;
2F=\begin{bmatrix}
     f_1\\
     .  \\
     .\\
     .\\
     f_{L_c}\\
     f_1^*\\
     .\\
     .\\
     .\\
     f_{L_c}^*
    \end{bmatrix}\,.
\end{equation}

Out of equilibrium it is straightforward to generalize Eq.~(\ref{condens}) to an equation along the contour: 
\begin{equation}\label{condens2}
 \hat{G}^{-1}\bullet \hat{A}=\hat{G}^{'-1}\bullet \hat{A}'+\hat{F}\,,
\end{equation}
where the ingredients are now contour functions. Again, the symbol $\bullet$ represents 
matrix multiplication in space and time integration along the contour.
We further simplify this expression by multiplying both sides of it 
by $\hat{G}$  from the left. This leads to 
\begin{equation}\label{condens3}
 \hat{A}=\hat{G} \bullet \hat{G}^{'-1}\bullet \hat{A}'+\hat{G} \bullet \hat{F}\,,
\end{equation}
where we have
used the fact that $\hat{G}\bullet \hat{G}^{-1}=\mathbb{1}$.
Via the CPT equation (\ref{ne-cpt-equation}) one can further derive the expression   
\begin{equation}
 \hat{G} \bullet \hat{G}^{'-1}=\mathbb{1}+\hat{G}\bullet \hat{V}\,. 
\end{equation}
After substituting into Eq.~(\ref{condens3}), one finally gets the following equation for the 
 condensate including CPT correction
\begin{equation}
 \hat{A}= \hat{A}'+\hat{G} \bullet \hat{V}\bullet \hat{A}'+\hat{G}\bullet \hat{F}\,.
\end{equation}
We rewrite this equation by expressing the contour integration explicitly as
\begin{equation}
 \hat{A}(z)=\hat{A}'(z)+\int_c d\bar{z}\, \hat{G}(z,\bar{z}) 
 \Big(V \hat{A}'(\bar{z}) +\hat{F}(\bar{z})\Big) \,,
\end{equation}
where $z,\bar{z}$ are contour variables (see Fig. \ref{contour}). By employing Langreth theorem \cite{Langreth} one can break down the contour 
integrations into contributions on the real and imaginary time axes. For the condensate on the real time branch of the contour we get 
 \begin{equation}
 \label{condensate}
 \begin{split}
A(t)=A'(t)&+\int_0^t d\bar{t}\, G^R(t,\bar{t}) \Big(V A'(\bar{t})+F(\bar{t})\Big)\\
&+\int_0^{-i\beta} d\bar{\tau}\, G^\rceil(t,\bar{\tau}) \Big(V A'(\bar{\tau})+F(\bar{\tau})\Big) \, ,
\end{split}
 \end{equation}
where we have used $G^R(t,t')=\theta(t,t') \Big(G^>(t,t')-G^<(t,t')\Big)$. Similarly for the condensate on the Matsubara branch we derive
\begin{equation}\label{cond_eq}
A(\tau)=A'(\tau)+\int_0^{-i\beta} d\bar{\tau}\,G^M(\tau,\bar{\tau})\Big(V A'(\bar{\tau})+F(\bar{\tau})\Big)\,.
\end{equation}
We note from Eq.~(\ref{condensate}) that, in order  to evaluate $A(t)$ within CPT, the mixing Green's function $G^\rceil$ and retarded Green's function $G^R$  
have to be determined first.  It is crucial to employ high-order numerical integration schemes to accurately 
simulate up to  long times. We refer to Appendix. (\ref{cpt-numeric}) for more details. 
\end{subsection}

\begin{subsection}{Magnetization}


 The time dependent magnetization is obtained as 
 \begin{equation}
  S^z(t)=\frac{1}{L}\sum_q \sum_{R=1}^{L_c} \, \big\langle \;a_{R,q}^\dag(t)\,a_{R,q}(t) - \frac{1}{2}\;\big\rangle\,, 
 \end{equation}
where $L=N_cL_c$ is the total size of the lattice. $\langle a_{R,q}^\dag(t)\, a_{R,q}(t)\rangle$ can be extracted from the lesser component of the Green's function within CPT
\begin{equation}
 \langle a_{R,q}^\dag(t)\,a_{R,q}(t)\rangle= i \,G_{RR,q}^<(t,t)\,.
\end{equation}
Adding the contribution from the condensate, the final expression for the magnetization is
\begin{eqnarray*}
 S^z(t)&=&\frac{1}{L}\sum_q \sum_R\, i \,G_{RR,q}^<(t,t)\\
 && +\frac{1}{L_c}\sum_{R=1}^{L_c} \,\bigg(\langle \,
 a_R^\dag(t)\,\rangle\; \langle \,a_R(t)\,\rangle-\frac{1}{2}\bigg)\,,
\end{eqnarray*}
 where $\langle a_R^\dag(t)\rangle$ and $ \langle a_R(t)\rangle$ are elements of the vector $A(t)$, see 
 Eq.~(\ref{condensate}).
\\
For a finite lattice with open boundary conditions, translation symmetry is lost and therefore the magnetization is position dependent, more pronounced close to the boundaries. 
\end{subsection}
\end{section}


\begin{section}{Results} \label{Results}
In the following we apply the technique  discussed in the previous section for both  equilibrium and nonequilibrium situations. Moreover, by comparing the results with exact ones in one dimension we  asses the accuracy of the method. 

\begin{subsection}{Equilibrium results}
 Before applying the technique out of equilibrium we investigate its ability to describe the system already in  equilibrium. This is actually a necessary step since the present non-equilibrium protocol assumes that the system is prepared as the ground state of an initial Hamiltonian and is then evolved with a different  Hamiltonian. 
Therefore, an accurate equilibrium state is a prerequisite for getting a sensible after-quench dynamics. 
 The CPT method can work directly in the thermodynamic limit,  however, in order to compare with exact results  in one dimension, we shall consider a finite system with linear size $L=8$ with open boundary conditions. In the CPT method the procedure is thus to divide the system into 
 two parts, $A$ and $B$, each one with size $L_c=4$, and then treat the inter-cluster term perturbatively, see Fig. \ref{lattice}. 
 
 \begin{figure}[t!] 
    \centering
    \includegraphics[scale=0.7] {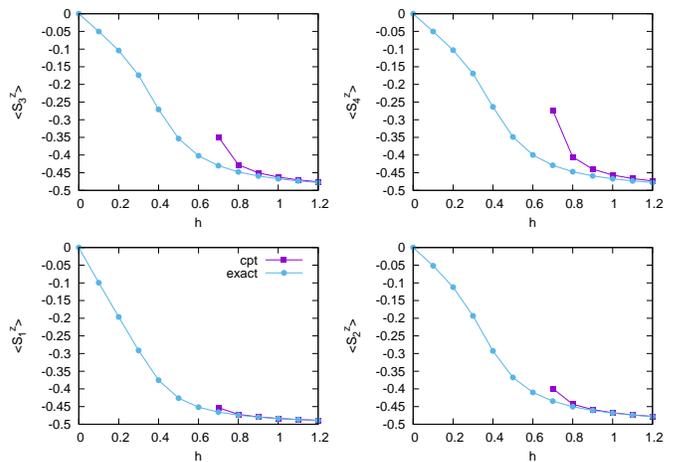}
    \caption{Magnetization along the $z$ direction versus magnetic field for different sites on a lattice of size $L=8$ with open boundary condition. Cluster size in CPT is $L_c=4$. Exact 
    results are also being reported for comparison.}
    \label{cpt-eq}
\end{figure}
We first set the anomalous term to zero in the cluster Hamiltonian, i.e, $\Delta=0$ in Eq.~(\ref{anomalous}).
In Fig.~\ref{cpt-eq} we display  the  magnetization parallel to the magnetic field, $\langle S^z\rangle $, for sites $i=1$ to $i=4$ compared with exact 
result. As we see CPT works well for large values of magnetic field and reproduces results close to exact ones. By contrast,  upon decreasing $h$ the accuracy 
decreases. Standard CPT totally fails close to the mean-field critical field ($h_c=0.7$). Therefore the standard CPT is unable to correctly describe the physics for $h< 0.7$. 

This kind of instability is well known in approaches based on the bosonic Bogoliubov approximation, such as the spin-wave approximation. The Green's function for free 
bosons ($U=0$ in the Hamiltonian of Eq. (\ref{hamilt-boson})) has two poles at $z=\pm \sqrt{h^2-\frac{J^2}{4}}$. It is clear that for $h<J/2$ the poles move to the imaginary axis, a clear signal of an instability. 
\begin{figure}[t!]
    \centering
    \includegraphics [scale=0.7]{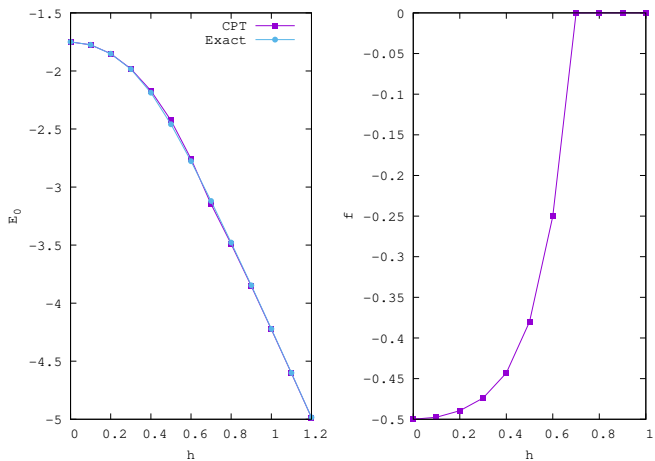}
    \caption{Left panel: Ground state energy from CPT compared with the exact value for lattice of size $L=8$ with open boundary condition. Right panel: $f=J\langle S_{1B}^x \rangle=J\langle S_{4A}^x \rangle$}
    \label{en-cpt}
\end{figure}
The same explanation  applies to the interacting Hamiltonian Eq. (\ref{hamilt-boson}) and to the instability seen in Fig. \ref{cpt-eq}. The poles of the Green's function 
become complex for small values of magnetic field, i.e. for $h < 0.7$. This is the region where the hard-core constraint of the bosons becomes important and the standard CPT fails to satisfy 
this condition. We control the location of the poles by adding the variational term $\Delta$ in Eq. (\ref{anomalous}) to the cluster Hamiltonian, which explicitly breaks the $Z_2$ symmetry and induces the spontaneous breaking of such symmetry at low fields. After finding 
self-consistently the optimum value for the variational parameters we compute the CPT corrections as explained in the previous section.

In Fig. \ref{en-cpt} on the left panel we show the result for the ground state energy compared with the exact one. 
The agreement is quite good
 in the whole range of magnetic fields. On the other hand, it is well known that the energy is a quantity that is not much sensitive to perturbations, so one could argue that this agreement is not significative.
On the other hand, the right panel shows the value of the 
variational parameter $f=J\langle S_{1B}^x \rangle=J\langle S_{4A}^x \rangle$. 
This quantity shows
a phase transition at $h_c=0.7$, below which $\langle S^x \rangle$ acquires a finite value. Strictly speaking such a phase transition should not occur on a finite size system, where $\langle S^x \rangle$ must be zero by symmetry, so its emergence is a spurious results that derives from the variational scheme. In the thermodynamic limit the transition does instead occur, although the critical field is known to be $h_c=0.5$. 
Nevertheless, by increasing the length $L_c$  of the cluster up to $L_c\approx 16$ we observe a decrease of $h_c$ to values close to the exact value.
For the time dependent calculation and for our benchmark, however, we have to stick to smaller values of $L_c\approx 4$.
Keeping in mind this caveat, let us turn to compare other physical observable different from $S^x$.

In Fig. \ref{occu-cpt} we report the $z$ magnetization on the sites $1,2,3,4$ compared with the exact value. As we see the comparison is quite satisfactory.  At site $i=1$ and for the whole range of magnetic fields, CPT results are very close to exact ones especially in the instability region $h\le 0.5$. Around $h=0.7$ the 
results are less close to the exact ones mainly for the site $i=4$ at the edge of the system.

\begin{figure}[t!] 
    \centering
    \includegraphics[scale=0.7] {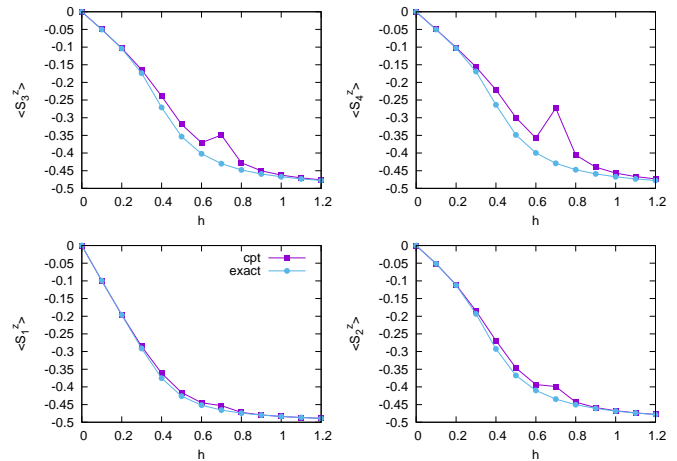}
    \caption{Magnetization of different sites versus magnetic field for a lattice of size $L=8$ with open boundary condition. The cluster size in CPT is $L_c=4$. Exact 
    results are also reported for comparison}
    \label{occu-cpt}
\end{figure}
It is worth mentioning that the hard-core constraint implies the following relation between expectation values:
\begin{equation}
 \langle a_i^\dag a_i\rangle+\langle a_i a_i^\dag\rangle=1\,.
\end{equation}
We found that within the present self-consistent CPT the expectation value of the above expression slightly deviates from one by about $10^{-3}$ on the average, with a  maximum of the order of $10^{-2}$ at $h=0.7$ and for the sites at the edge of the supercell, as shown by the kink around $h=0.7$ in Fig.~\ref{occu-cpt}. This is due to the fact that treating inter-cluster terms perturbatively 
violates the constraint within CPT, mainly for the edge sites.   
 \end{subsection}
 
 \begin{subsection}{Non-equilibrium results}
In this section we present results for the real time dynamics of the Ising model within the variational cluster perturbation approach introduced above. To drive the system out of equilibrium we 
proceed as follows.
 We prepare the system at equilibrium for $t_0<0$ as the ground state 
of \eeqref{hamilt}
with magnetic field $h_0$ and then we suddenly change the magnetic field to 
a different value $h$. As in equilibrium we use variational NE-CPT in a self-consistent way as described in section (\ref{Method}). After finding the time dependent 
variational parameters for each time step in the mean-field approximation, we calculate Green's function and condensate within CPT as described in Sec.~\ref{Method}.  

To benchmark this idea for the non-equilibrium case we display in Fig. \ref{L8-h0.2-h1.2} the real time dynamics of magnetization for different 
sites on a lattice of size $L=8$ with open boundary condition at zero temperature. We compare results obtained exactly with results within CPT 
for a cluster size of $L_c=4$. We have reported the dynamics for the case of relatively large quench, from $h_0=0.2$ to $h=1.2$, for which the field crosses the phase transition. 
As we can see, NE-CPT provides 
quite good results for the magnetization compared to the exact one except for the edge sites where the hopping to the next supercell is treated perturbatively. At the beginning of the dynamics 
 NE-CPT is very accurate and the deviation builds up as time progresses.
 
 \begin{figure}[t!]
    \includegraphics[scale=0.7]{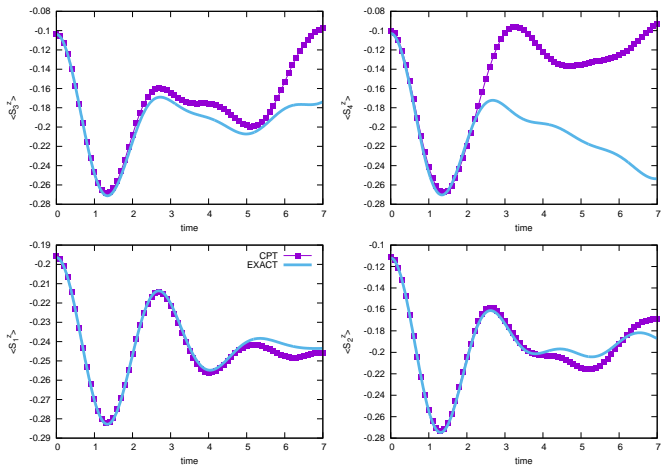}
    \caption{Time dependence of the magnetization in the $z$ direction for different sites on a lattice of size $L=8$. The cluster size in CPT is $L_c=4$. The magnetic 
    field has been suddenly changed from $h_0=0.2$ to $h=1.2$. The exact dynamics is shown for comparison.}
    \label{L8-h0.2-h1.2}
\end{figure}
 We have investigated different types of quenches and the behavior is qualitatively the same:  the dynamics remains close to the exact one at short times and starts deviating at later times. As mentioned, the largest deviations are found at the edge sites. 
 
In Fig. \ref{1d-infinite} we report 
 NE-CPT results for  the magnetization dynamics after the quench  
for an infinite lattice. We display results obtained 
for different cluster sizes and different types of quenches.
For quenches 
into the ordered phase (see lower left and right panels),  NE-CPT for a cluster of $L_c=6$ provides quite accurate results for the 
 magnetization dynamics up to $t\approx 7$ (remember, that time is in unit of $1/J$).
 For a larger quench from $h_0=1.2$ to $h=0.4$, crossing the transition point,  NE-CPT is able to 
reproduce  the dynamics only up to a shorter value of the time  $t\approx 5$ (see upper left panel). For quenches within the disordered 
phase (quench from $h_0=1.2$ to $h=1.6$) NE-CPT results  show only a slight deviation
($\lesssim 10^{-3}$) 
 from the exact one, however with some small oscillations (see upper right panel). 
Overall, NE-CPT results 
 for an infinite system  systematically improve  by increasing 
cluster size $L_c$.  Already for  $L_c\approx 6$ they reproduce quite accurate results for the thermodynamic limit of 
a very long chain ($L=400$)
up to $t\approx7$.  \\
Finally we report the real time quench dynamics of the magnetization for the two dimensional Ising Model, which is not exactly solvable. In $D=2$ the transverse field Ising model at zero temperature has an 
equilibrium phase transition at $h_c\approx 1.6$\cite{Friedman,Elliott,Oitmaa}. 
\begin{figure}[t!]
   \includegraphics[scale=0.7]{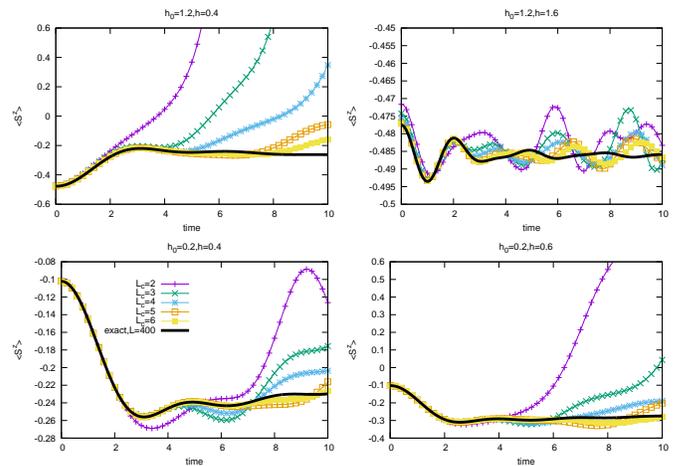}
   \caption{Time dependence of the magnetization 
for an infinite ising chain evaluated within NE-CPT with differnt cluster sizes $L_c$ compared with exact results for a chain of length $L=400$.
 }
   \label{1d-infinite}
\end{figure}
Within the self-consistent variational CPT at equilibrium we get instead $h_c\approx 1.9$, for the small $2\times 2$ clusters we are considering.
 We note that the accuracy of CPT  improves systematically by increasing the cluster size. We show 
results for different types of quenches which are obtained either within a disordered or a ordered phase or a quench which crosses the critical field. 
Here we compare  Lanczos exact results obtained for three different lattice sizes with periodic 
boundary condition to  NE-CPT results using  clusters of size $L_c=2\times2$.
 We note that the largest  size we can reach  to perform real time quench 
dynamics 
within Lanczos
at zero temperature is $L=4\times 4$.

For a small quench in the ordered phase, $h_0=0.2$ to $h=0.4$
(Fig. \ref{2d-infinite} lower left-panel)   
 NE-CPT gives quite accurate results, as compared with  Lanczos exact results. 
We observe that magnetization dynamics within NE-CPT is quite close to the best of the Lanczos for times up to $t_{max}=10$. We further note 
that in this case the Lanczos results already show convergence 
as a function of system size
so that
 they can be considered as a good approximation to the thermodynamic limit for these values of the parameters. 
For  a larger quench but still in the ordered phase, i.e. $h_0=1.2$ to $h=0.4$ (top left-panel 
of Fig. \ref{2d-infinite})
the NE-CPT results compare well with Lanczos results up to  $t_{max}\approx6$.
 For quenches with large magnetic 
fields, i.e. into the disordered phase from $h_0=2.0$ to $h=2.5$ 
(top right-panel in Fig. \ref{2d-infinite}) 
the Lanczos results have not converged yet, so a comparison is difficult to assess.
Nevertheless, 
 the NE-CPT results
 quantitatively agrees with the largest Lanczos system up to $t\approx 2$, and agrees
qualitatively, i.e. displays similar oscillations, also for larger times.
 The lower right-panel in Fig. \ref{2d-infinite} shows the 
dynamics for a large magnetic quench that crosses the critical point, i.e. from $h_0=0.2$ to $h=2.0$. In this case  the NE-CPT seems not to be accurate and is able to produce reliable dynamics only up to $t_{max}\approx2$. We note that in all cases  the magnetization 
stays within its physical values, $\left|S^z\right|  \le \frac{1}{2}$, except for a quench across the critical point. 
With  cluster sizes of $L_c=2\times 2$  results are already promising in two dimensions, as long as one restricts to intermediate times. Furthermore, as in the $D=1$ case, NE-CPT results can be improved 
 by systematically increasing the cluster size.
\begin{figure}[t!]
   \includegraphics[scale=0.7]{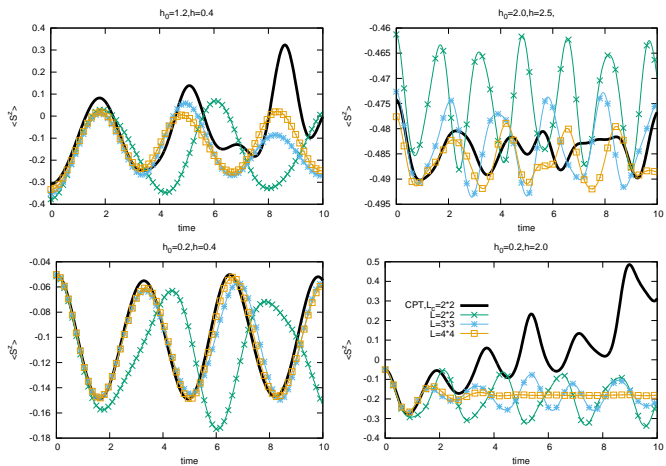}
   \caption{Dynamics for the magnetization compared to Lanczos results for different types of quenches in two dimension. NE-CPT calculation is for an infinite lattice with the cluster 
   size of $L_c=2\times 2$.}
   \label{2d-infinite}
\end{figure}

\end{subsection}

\end{section}

\begin{section}{Summary}\label{Summary}
We have introduced a  variational formulatiuon of cluster perturbation theory (CPT)  to investigate the quantum Ising model in and out of equilibrium at zero temperature. 
We find that plain CPT in equilibrium can describe accurately the system in the disordered phase, $h>h_c$, but looses accuracy while approaching the critical field $h_c$ and finally breaks down 
in the ordered phase, $h<h_c$. To describe the system in the broken symmetry region we developed a variational implementation of CPT, whereby  an anomalous term is added to the cluster Hamiltonian
and subtracted perturbatively. This parameter 
is then optimized 
 within a self-consistent framework.
We find a good agreement with exact results in the equilibrium case, for example concerning the
 magnetisation parallel to the magnetic field and the 
ground state energy.

Out of equilibrium the time dependent variational parameter are determined self-consistently 
for each time step.
 We find 
that this variational NE-CPT provides very accurate results for the short and intermediate time dynamics while getting inaccurate for longer times. Specifically in one dimension, comparing results of this 
NE-CPT approximation 
with exact calculations shows that clusters of size $L_c=6$ provide quite accurate description of the dynamics up to $t_{max}\approx 7$ for quenches within the ordered or disordered phases. When the critical point is crossed, the accuracy is limited to  shorter times $t_{max}\approx 3$. A similar trend emerges also in two dimensions.
Here, there is no  exact solution to be compared with, so that we resort to finite-size Lanczos diagonalization to benchmark the method.
One should notice, however, that  NE-CPT can directly provide results in the thermodynamic limit. We highlight that the accuracy of NE-CPT can be systematically pushed to longer time by increasing the 
cluster size, at least up to the largest sizes still reachable by  Lanczos time evolution.

This variational approach shall be considered as a first step to study hard-core bosons in and out of equilibrium by a variational CPT method. This idea could be improved and 
getting more elaborate by considering more variational terms 
and/or formulating it within the self-energy functional theory\cite{Hofmann}.

\end{section}

\begin{section}{Acknowledgment}
We would like to thank M. Nuss, M. Aichhorn and A. Dorda for fruitful discussions.
This work was supported by funds awarded by the Friuli Venezia Giulia autonomous Region Operational Program of the European Social 
Fund 2007/2013, Project DIANET-Danube Initiative and Alps Adriatic Network, CUP G93J12000220009. The work was also 
partially supported by the Austrian Science Fund (FWF): P26508, Y746, and  NaWi Graz.

\end{section}


\begin{appendices}
\section{Numerical solution of CPT equation}
\label{cpt-numeric}
\subsection{Time propagation}
 For spatially inhomogeneous systems, the computational limits are set by the memory requirement for saving  big matrices in two time and spatial degrees of freedom. 
 Considering $N_K$ time steps on the Keldysh and $N_M$ on the Matsubara for a lattice of size $L$ the required memory to save the Green's function is
 \begin{equation}
  (2*L*(2*N_K+N_M))^2*16~~~ bytes
 \end{equation}
Therefore, for a $N_K=1000, t_{max}=10$, $N_M=2000, \beta=10$, $L=8$ the required memory is $61$ Gigabytes. This example shows that reaching large time or large system sizes ($t\gtrsim 10 J, L\gtrsim 6$) is prohibitive 
since the memory requirement is beyond the capabilities of standard available computational resources. 
Another issue is the inversion of the huge matrix in the CPT equation to get the lattice Green's function. When the matrix size increases the inversion process takes 
longer time and also the numerical error will increase. \\
Based on the above facts one has to design a way to avoid matrix inversion and the storage of huge matrices to finally be able to reach longer times for the dynamics of the system. 

\subsection{Procedure to propagate the CPT equation in time}
 In this section, we summarize the procedure to numerically solve the CPT equation by gradually progressing in time in order to avoid inversion and storage of big matrices. Since this equation is of the Kadanoff-Baym type, there is a great deal of literature on the subject (see, e.g. \cite{Bonitz}).
For example, the method is used in 
time dependent dynamical mean-field theory (TDMFT) \cite{Aoki,Tran}, where, however  
the system is typically translationally invariant. 
In our case, where we deal with finite systems, we have
 to consider spatial degrees of freedom and, accordingly, solve a corresponding set of equations for inhomogeneous system.

Here, we roughly follow  the treatment of Ref.~\cite{Tran}, see also ~\cite{Bonitz}. Here, in addition, we consider  the case of an inhomogenous system. 
 The CPT equation for the Green's function $ \hat{G}$ of the physical system is
 \begin{equation}
 \label{cpt1}
  \hat{G}=\hat{G_0}+\hat{G_0} \bullet \hat{V} \bullet \hat{G}
 \end{equation}
 Where $\hat{G_0}$ is the cluster Green's function and $\hat{V}=V \bigotimes \mathbb{1}$ is the inter-cluster term. 
 By introducing $\hat{K}=\hat{G_0}\bullet \hat{V}$ we rewrite the CPT equation as 
 \begin{equation}
  \hat{G}=\hat{G_0}+\hat{K} \bullet \hat{G}
 \end{equation}
 and after writing the contour integration explicitly we have
 \begin{equation}
 \hat{G}(z,z')=\hat{G}_0(z,z')+\int_{\cal {C}} dz_1 \hat{K}(z,z_1)\hat{G}(z_1,z')
\end{equation}

 Using Langreth theorem \cite{Langreth} we can write the integral equation for the components of the Green's function on the contour. We obtain the following equations:
  \begin{equation}
 \begin{split}
 G^<=G_0^<+K^<\cdot G^A+K^R\cdot G^<+K^\rceil*G^\lceil \\
 G^>=G_0^>+K^>\cdot G^A+K^R\cdot G^>+K^\rceil*G^\lceil\\
 G^R=G_0^R+K^R\cdot G^R\\
 G^A=G_0^A+K^A\cdot G^A\\
 G^\rceil=G_0+K^R\cdot G^\rceil+K^\rceil*G^M\\
 G^\lceil=G_0^\lceil+K^\lceil \cdot G^A+K^M*G^\lceil\\
 G^M=G_0^M+K^M*G^M
\end{split}
\end{equation}
where $.$ means integration over real time and $*$ over imaginary time (Matsubara branch).\\
We are interested in the magnetization which can be calculated from the lesser ($G^<$)
 Green's function. To Solve the equation for $G^<$ first we need to calculate $G^A$ and $G^\lceil$ within CPT. 
Furthermore to determine $G^\lceil$ we need to evaluate the Matsubara Green's function $G^M$ which can be calculated  with equilibrium techniques. 
The cluster Green's function $G_0^\alpha$, ($\alpha=<,R,A,>,\lceil,\rceil$) also should be calculated for an affordable cluster size. \\
It is also  worth to mention that due to the presence of  
anomalous terms like $a_i a_j$ and $a_i^\dag a_j^\dag$ in the Hamiltonian of Eq. (\ref{hamilt-boson}) the structure of the Green's function matrix 
also should include anomalous Green's functions in order to satisfy the correct equation of motion. Therefore the $G^\alpha$, ($\alpha=<,R,A,>,\lceil,\rceil$), is a matrix in itself 
with the following Nambu structure
\begin{equation}
 G^\alpha=
 \begin{bmatrix}
  g^\alpha  &  f^\alpha\\
  f^{\alpha \dag} & k^\alpha
 \end{bmatrix}
\end{equation}

The definition of the Green's functions are as follow:\\
1) Advanced Green function:
\begin{equation}
\begin{split}
g_{i,j}^A(t,t')=i\theta(t'-t)[\langle a_i(t)a_j^\dag(t')\rangle-\langle a_j^\dag(t')a_i(t)\rangle]\\
k_{i,j}^A(t,t')=i\theta(t'-t)[\langle a_i^\dag(t)a_j(t')\rangle-\langle a_j(t') a_i^\dag(t)\rangle]\\
f_{i,j}^A(t,t')=i\theta(t'-t)[\langle a_i(t)a_j(t')\rangle-\langle a_j(t')a_i(t)\rangle]\\
f_{i,j}^{\dag A}(t,t')=i\theta(t'-t)[\langle a_i^\dag(t)a_j^\dag(t')\rangle-\langle a_j^\dag(t')a_i^\dag(t)\rangle]\\
\end{split}
\end{equation}
2) Retarded Green's function:
\begin{equation}
\begin{split}
g_{i,j}^R(t,t')=-i\theta(t-t')[\langle a_i(t)a_j^\dag(t')\rangle-\langle a_j^\dag(t')a_i(t)\rangle]\\
k_{i,j}^R(t,t')=-i\theta(t-t')[\langle a_i^\dag(t)a_j(t')\rangle-\langle a_j(t') a_i^\dag(t)\rangle]\\
f_{i,j}^R(t,t')=-i\theta(t-t')[\langle a_i(t)a_j(t')\rangle-\langle a_j(t')a_i(t)\rangle]\\
f_{i,j}^{\dag R}(t,t')=-i\theta(t-t')[\langle a_i^\dag(t)a_j^\dag(t')\rangle-\langle a_j^\dag(t')a_i^\dag(t)\rangle]\\
\end{split}
\end{equation}
There is relation between retarded and advanced green's function: $G^R(t,t')=G^{A\dag}(t',t)$.\\
3) Lesser Green's function
\begin{equation}
\begin{split}
 g_{i,j}^<(t,t')=-i\langle c_j(t')^\dag c_i(t)\rangle \\
 f_{i,j}^<(t,t')=-i\langle c_j(t') c_i(t)\rangle \\
 f_{i,j}^{<\dag}(t,t')=-i\langle c_j(t')^\dag c_i(t)^\dag\rangle \\
 k_{i,j}^<(t,t')=-i\langle c_j(t') c_i(t)^\dag\rangle 
\end{split}
\end{equation}
4) Mixing Green's function:
\begin{equation}
 \begin{split}
  g^\lceil_{i,j}(\tau,t)=-i\langle c_i(\tau)c_j^\dag(t)\rangle \\
  f^\lceil_{i,j}(\tau,t)=-i\langle c_i(\tau)c_j(t)\rangle \\
  f^{\lceil \dag}_{i,j}(\tau,t)=-i\langle c_i^\dag(\tau)c_j^\dag(t)\rangle \\
  k^\lceil_{i,j}(\tau,t)=-i\langle c_i^\dag(\tau)c_j(t)\rangle \\
  g^\rceil_{i,j}(t,\tau)=-i\langle c_j(\tau)^\dag c_i(t)\rangle \\
  f^\rceil_{i,j}(t,\tau)=-i\langle c_j(\tau) c_i(t) \rangle \\
  f^{\rceil \dag}_{i,j}(t,\tau)=-i\langle c_j(\tau)^\dag c_i(t)^\dag \rangle \\
  k^\rceil_{i,j}(t,\tau)=-i\langle c_j(\tau) c_i(t)^\dag \rangle \\
 \end{split}
\end{equation}
If we write the integral equation for $G^A$ we get (omitting the spatial indexes):

\begin{equation}
 \begin{split}
  G^A(t,t')=G_0^A(t,t')+\int K^A(t,\bar{t})G^A(\bar{t},t') d\bar{t} \\
  G^A(t,t')=G_0^A(t,t')+\int_t^{t'} K^A(t,\bar{t})G^A(\bar{t},t') d\bar{t}\\
\end{split}
\end{equation}
To perform the integration, we discretize the time with equal spacing
\begin{equation}
\begin{split}
 t_i=i\times \Delta t+t_0,~~ \Delta t=\frac{t_{max}-t_0}{N_K-1}, ~~(i=0,1,...,N_K-1)\\
 \tau_i=i\times \Delta \tau+t_0, ~~\Delta \tau=\frac{-i\beta-t_0}{N_M-1},~~(i=0,1,...,N_M-1)
\end{split}
\end{equation}
where $N_K$ and $N_M$ are the number of time points on the real and imaginary branch respectively. After approximating the integral by the trapezoid 
rule 
\begin{equation}
\begin{split}
\int_{t_a}^{t_b} f(x)dx\approx\Delta t\sum_{i=0}^{N-1}\omega_i f(x_i),~~~ \Delta t=\frac{t_b-t_a}{N-1} \\
w_i=\left\{
\begin{array}{cc}
 1/2 & i=0,N-1\\
 1 &  1\le i\le N-2
\end{array}
\right. 
\end{split}
\end{equation}
we get 

\begin{equation}
 \begin{split}
  G^A(t_m,t'_n)\approx G_0^A(t_m,t'_n)+\Delta \bar{t}\sum_{i=m}^n w_i K^A(t_m,\bar{t_i})G^A(\bar{t_i},t'_n)  \\
  G^A(t_m,t'_n)\approx G_0^A(t_m,t'_n)+\Delta \bar{t}\sum_{i=m+1}^n w_i K^A(t_m,\bar{t_i})G^A(\bar{t_i},t'_n)  \\
\end{split}
\end{equation}
where $K^A(t_m,t_m)=0$ is used. In the equation for $G^A$ it is not possible to gradually propagate in the direction of time since $G^A(t_m,*)$ depends on later times $m+1,...n$, in other
words this equation is not of the Volterra type  \cite{Brunner} where the causal structure is evident from the limits of the integral. To get a
Volterra type of equation for the $G^A$ we have to use another form of the CPT equation:

\begin{equation}
\label{cpt2}
  \hat{G}=\hat{G_0}+\hat{G} \bullet \hat{V} \bullet \hat{G_0}
 \end{equation}
where now $\hat{K}=\hat{V}\bullet \hat{G_0}$.
Proceeding in the same way as above we derive the following equation in discretized time for the advanced Green's function
\begin{equation}
\boxed{
G^A(t_m,t'_n)\approx G_0^A(t_m,t'_n)+\Delta \bar{t} \sum_{i=m+1}^{n-1} w_i G^A(t_m,\bar{t_i})K^A(\bar{t_i},t'_n)  
}
\end{equation}
We now  can gradually proceed in the second index $t'_n$ for a fixed $t_m$.\\
Similarly for the retarded Green's function we get

\begin{equation}
\boxed{
\begin{split}
G^R(t_{m1},t'_{m2})\approx G_0^A(t_{m1},t'_{m2})+\\
\Delta \bar{t} \sum_{i=m2+1}^{m1-1} w_i K^R(t_{m1},\bar{t_i})G^R(\bar{t_i},t'_{m2})  
\end{split}
}
\end{equation}
where we can progress in time by incrementing $t_{m1}$ for a fixed $t'_{m2}$.\\
For the mixed Green's function if we use the CPT Eq. (\ref{cpt2}) we obtain the following integral equation:
\begin{equation}
\begin{split}
 G^\lceil(\tau,t)\approx G_0^\lceil(\tau,t)+\int_0^{t'} K^\lceil(\tau,\bar{t}) G^A(\bar{t},t) d\bar{t}+\\
 \int_0^{-i\beta} K^M(\tau,\bar{\tau})G^\lceil(\bar{\tau},t) d\bar{\tau}
 \end{split}
\end{equation}
Since in the convolution including $G^\lceil$ the integration over $\tau$ is on the whole Matsubara branch, it is not possible to gradually proceed in time. So the way out is to choose the 
other CPT Eq. (\ref{cpt2}) to end up in a Volterra type equation
\begin{equation}
 \boxed{
 \begin{split}
 G^\lceil(\tau_{m1},t_{m2})\approx G_0^\lceil(\tau_{m1},t_{m2})+\\
 \Delta t\sum_{i=0}^{m2-1} w_i G^\lceil(\tau_{m1},t_{i}) K^A(t_{i},t_{m2})\\
 +\Delta\tau \sum_{i=0}^{N_M-1} w_i G^M(\tau_{m1},\tau_i) K^\lceil(\tau_i,t_{m2}) \;.
 \end{split}
 }
\end{equation}
Here one 
 can  proceed in $t_{m2}$ for a fixed $\tau_{m1}$. The full information of the Matsubara Green's function on the imaginary axis is necessary to calculate the mixing Green's function. 
This can be done by equilibrium techniques, see appendix \ref{equi-green}.\\
Finally for the lesser greens function we have:

\begin{equation}
 \boxed{
 \begin{split}
 G^<(t_{m1},t_{m2})\approx G_0^<(t_{m1},t_{m2})+\\
 \Delta t\sum_{i=0}^{m2-1} w_i K^<(t_{m1},t_{i}) G^A(t_{i},t_{m2})\\
 +\Delta\tau \sum_{i=0}^{N_M-1} w_i K^\rceil(t_{m1},\tau_i) G^\lceil(\tau_i,t_{m2})\\
 +\Delta t\sum_{i=0}^{m1-1} w_i K^R(t_{m1},t_i)G^<(t_i,t_{m2})
 \end{split}
 }
\end{equation}

 \section{Calculating equilibrium Green's function $G^M$}
 \label{equi-green}
The CPT equation for $G^M$ is not of the Volterra type, so it is not possible to gradually proceed along the Matsubara axes. 
Fortunately due to the time translation invariance of the Green's function, $G^M$ only depend on the time difference and so one can use Fourier transformation to go over to 
the frequency representation. In this way one still has to do inversion process to get CPT Green's function but in this way the dimension reduces to the size of the lattice.\\
The Fourier transformations between the Green's functions are as follow:
\begin{equation}
\label{FT}
\begin{split}
 G(\tau)=\frac{1}{\beta} \sum_{n=-\infty}^{\infty} G(i\omega_n) e^{-i \omega_n \tau}\\
 G(i\omega_n)=\int_0^\beta d\tau e^{i \omega_n \tau} G(\tau)
 \end{split}
\end{equation}
where $\tau \in [0,\beta]$. By using fast Fourier transformation (FFTW) the above transformation can be carried out  efficiently. When doing the inverse transformation 
we  truncate the number of Matsubara frequencies. Using $N$ points equally distributed among  positive and negative frequencies we get the approximation
\begin{equation}
\begin{split}
 G(\tau)\approx \frac{1}{\beta}\sum_{n=-N/2}^{N/2-1} G(i\omega_n) e^{-i \omega_n \tau}=DIFT[G(i\omega_n)] \\
 \omega_n=\frac{\pi}{\beta}2n
 \end{split}
\end{equation}
This scheme poorly  describes $G(\tau)$ due to missing contributions from the tail of $G(\i\omega)$ ($\omega_n\to \infty$). In practice, it is not possible to consider an infinite number of 
frequencies so one should calculate the tail correction directly. If we look at the asymptotic behavior ($\omega_n\to \infty$) for the non-interacting Green's function we realize
\begin{equation}
 G(i\omega_n) \sim -\frac{i}{\omega_n}
\end{equation}
The asymptotic tail of $G_A(\tau)$ can be readily calculated  by doing the Fourier transformation. For bosons we get:
\begin{equation}
 G_A(\tau)=-\frac{2}{\beta}\sum_{n=0}^\infty \frac{\sin(\omega_n \tau)}{\omega_n}=-\frac{1}{2}+\frac{\tau}{\beta}
\end{equation}
After a little algebra we can collect all contributions at high imaginary frequencies in the tail of the Green's function $G_T(\tau)$ and write:
\begin{equation}
\begin{split}
 G(\tau)=DIFT[G(i\omega_n)]+G_T(\tau)\\
 G_T(\tau)=-\frac{1}{2}+\frac{\tau}{\beta}+\frac{2\pi}{\beta}\sum_{n=0}^{N/2-1} \frac{\sin(\omega_n \tau)}{\omega_n}
 \end{split}
\end{equation}

\end{appendices}


\end{document}